\documentclass[%
 reprint,
superscriptaddress,
 amsmath,amssymb,
 aps,
 nourl,
]{revtex4-2}

\usepackage{mathrsfs}
\usepackage{xcolor}
\usepackage{graphicx}
\usepackage{dcolumn}
\usepackage{bm}

\usepackage[hidelinks]{hyperref}
\usepackage{braket}
\usepackage{dsfont}
\usepackage{tikz}
\usepackage{tikz-cd}

\newcommand{\beq}{\begin{equation}}
\newcommand{\eeq}{\end{equation}}

\newcommand{\Sch}[2]{\bigl\{#1,#2\bigr\}}

\begin{document}

\preprint{APS/123-QED}

\title{A UV-Finite Refinement of the Ryu–Takayanagi Relation in AdS$_3$/CFT$_2$}

\author{Philipp Dorau}
\email{philipp.dorau@uni-leipzig.de}
\affiliation{Institut f\"ur Theoretische Physik, Universit\"at Leipzig, Br\"uderstraße 16, 04103 Leipzig, Germany}

\author{Albert Much}
\email{amuch@sharjah.ac.ae}
\affiliation{Quantum Information Research Group, Department of Applied Physics and Astronomy, College of Sciences, University of Sharjah, P.O. Box 27272, Sharjah, United Arab Emirates}

\author{Leonardo Sangaletti}
\email{leonardo.sangaletti@edu.unige.it}
\affiliation{Dipartimento di Fisica, Università di Genova, Via Dodecaneso 33, 16146 Genova, Italy}

\author{Rainer Verch}
\email{rainer.verch@uni-leipzig.de}
\affiliation{Institut f\"ur Theoretische Physik, Universit\"at Leipzig, Br\"uderstraße 16, 04103 Leipzig, Germany}

\date{October 16, 2026}

\begin{abstract}
We establish a refined Ryu--Takayanagi (RT) relation in AdS$_3$/CFT$_2$ using \emph{relative entropy} as the central object, in place of the ultraviolet-divergent von Neumann entanglement entropy. Adapting Hollands' exact result for the chiral relative entropy to a diamond region, we express the boundary relative entropy between the vacuum and a coherent state as a Schwarzian functional, which the Fefferman--Graham dictionary identifies with the asymptotic data of a Ba\~nados geometry;  the rigidity of three-dimensional gravity promotes this boundary identification to the bulk. To linear order in the metric perturbation, the relative entropy then equals the variation of the RT geodesic length divided by $4G_N$. The construction rests only on the Bisognano--Wichmann/Borchers theorem and the holographic dictionary, giving a UV-finite, operator-algebraic counterpart to the RT relation.
\end{abstract}

\maketitle

\paragraph*{Introduction.}
The Ryu--Takayanagi (RT) proposal~\cite{Ryu:2006bv} states that, in a holographic conformal field theory (CFT), the entanglement entropy \(S_A\) of a spatial subregion \(A\) is given by the area of a  {static} codimension-two minimal surface \(\gamma_A\) in the anti--de Sitter (AdS) geometry, anchored at \(\partial A\) and homologous to \(A\), namely
\begin{equation}
  S_A \;=\; \frac{\mathrm{Area}(\gamma_A)}{4G_N}\,,
  \label{eq:RT-intro}
\end{equation}
where \(G_N\) denotes Newton's constant. In the special case of AdS$_3$/CFT$_2$, the surface \(\gamma_A\) is a spacelike geodesic, so \(\mathrm{Area}(\gamma_A)\) reduces to its length. The RT proposal is in close analogy with the Bekenstein--Hawking formula for black hole entropy~\cite{Ryu:2006bv,Nishioka:2009rev}; it and its covariant generalisations have been verified in a wide range of examples and applied broadly to quantum information, see, e.g.\ Refs.~\cite{Headrick:2010proof,Casini:2011kv,Lewkowycz:2013nqa,Faulkner:2013ana,Almheiri:2015}.

From the perspective of operator-algebraic quantum field theory (AQFT), however, the von Neumann entanglement entropy $S_A$ in the RT relation~\eqref{eq:RT-intro} is UV-divergent and regulator dependent: it is not a well-defined quantity for the type~III local algebras of QFT. Existing derivations of the proposal~\eqref{eq:RT-intro} and its refinements---via the replica trick~\cite{Lewkowycz:2013nqa} (see also Ref.~\cite{Jiang_2026}), quantum error correction~\cite{Almheiri:2015}, or linearised arguments around the vacuum geometry of AdS~\cite{Lashkari:2013,Faulkner:2013ica,Blanco:2013}---therefore work either with this divergent quantity or with subtracted, renormalised versions thereof.

By contrast, the \emph{relative entropy} \cite{Araki:1976, Uhlmann:1977relent, Witten_2018} is UV-finite and mathematically well-defined: it measures the distinguishability between two states on a given algebra and is tied to modular theory and horizon thermodynamics.

In this paper we follow the ideas of Refs.~\cite{KPV:2021ea,D_Angelo_2024,Dorau_2026}, where the relative entropy is related to a geometric area variation, and develop this proportionality concretely in the AdS$_3$/CFT$_2$ setting \footnote{In the context of AQFT, holography has been initially treated in Refs. \cite{Rehren:1999AlgebraicHolography, Rehren:2000LocalQObsAdSCFT, Duetsch:2003GFFAdSCFT}, see also Ref. \cite{Dappiaggi:2006HolographyAsymptoticallyFlat}.}.
Instead of assigning a finite entanglement entropy to a single state, we compare the vacuum $|\Omega\rangle$ with a family of chiral Virasoro coherent states $|\Psi_f\rangle = U(f)|\Omega\rangle$ \footnote{They are called Virasoro states because they are generated by elements of the Virasoro algebra, see Refs.~\cite{FEWSTER_2005,Hollands:2019}.} on a boundary interval $I$, and work with the Araki--Uhlmann relative entropy \cite{Araki:1976,Uhlmann:1977relent}, denoted by $S^I_{\mathrm{rel}}(\omega_0\|\omega_f)$, \footnote{Here $\omega_0$ and $\omega_f$ denote states on the algebra, and $\vert\Omega\rangle$ and $\vert\Psi_f\rangle$ are their corresponding vector representatives in the respective GNS representation.}. { The interval $I$ is the base of the causal diamond $C_I$ and coincides with the Ryu--Takayanagi region $A$; we use $I$ for the algebraic interval and $\gamma_A$ for its bulk geodesic, anchored at $\partial A=\partial I=\{\pm\ell\}$.}
To compute the relative entropy for the CFT on the boundary, we employ Hollands' exact formula, which is given in terms of a Schwarzian functional of the boundary reparametrisation~\cite{Hollands:2019}, see as well \cite{Panebianco_2020}. This Schwarzian fixes the expectation value of the chiral boundary stress tensor, which enters the Fefferman--Graham expansion~\cite{FeffermanGraham1985, FeffermanGraham2007} as the source for the bulk metric. {  In the setting of pure AdS$_3$ gravity, i.e. retaining only the stress-tensor data and neglecting the bulk matter sourced by the other excited boundary operators (cf.~\cite{martinec1998cft}), the rigidity of three-dimensional gravity then determines the dual geometry uniquely---the Ba\~nados spacetime corresponding to the coherent excitation $|\Psi_f\rangle$~\cite{Banados:1998, Roberts_2012}.}

\begin{figure}[h!]
    \centering
    \includegraphics[width=0.37\textwidth]{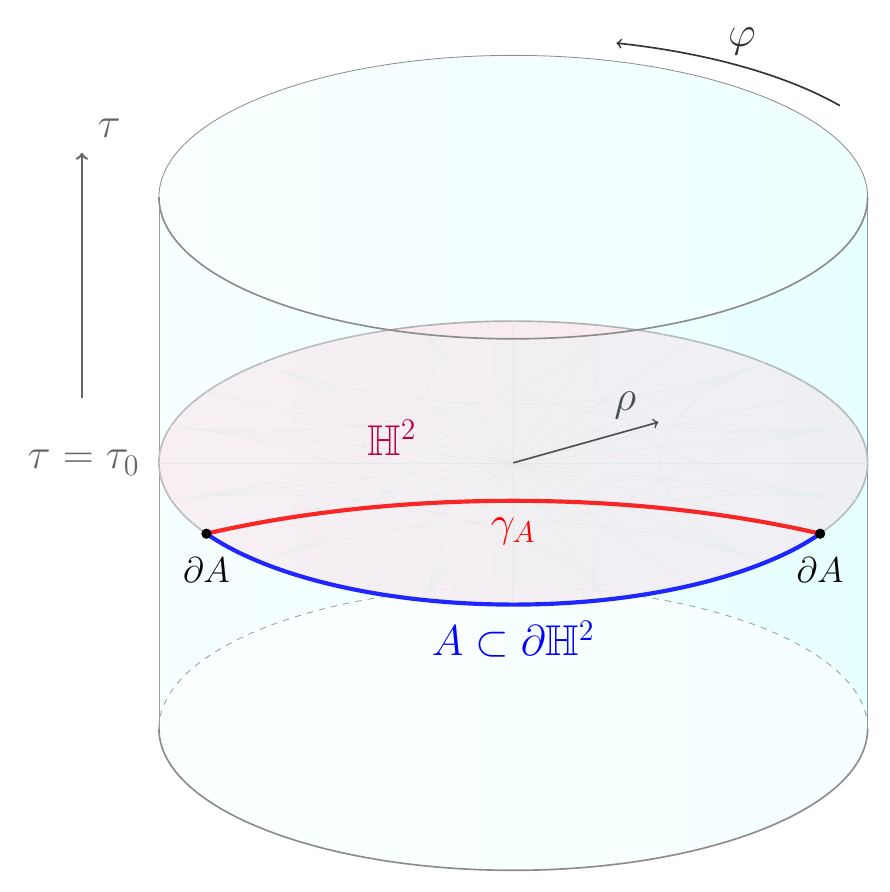}
    \caption{Schematic visualization of AdS$_3$ and its conformal boundary in global coordinates $(\tau,\rho,\varphi)$ (cf. \cite{Ryu:2006bv}). Each slice of constant $\tau = \tau_0$ corresponds to the hyperbolic plane $\mathbb{H}^2$. The segment $A$ represents a spatial subregion on the boundary $\partial\mathbb{H}^2$. The RT minimal surface $\gamma_A$ is the codimension-two bulk geodesic corresponding to $A$, satisfying $\partial\gamma_A=\partial A$.}
    \label{FigureAdS3}
\end{figure}

\paragraph*{Main Result.}To linear order in the deviation of the Ba\~nados geometry from the AdS$_3$ vacuum, the relative entropy of a coherent state $|\Psi_f\rangle$ in the chiral CFT$_2$ with respect to the vacuum $|\Omega\rangle$, restricted to an interval $I$ on the boundary (cf. Fig.~\ref{FigureAdS3}), \emph{equals} the variation of the Ryu--Takayanagi geodesic length divided by $4G_N$.  This establishes a rigorous, UV-finite RT relation, without any reference to the divergent von Neumann entanglement entropy. Concretely, we find that
\begin{equation}
S^I_{\mathrm{rel}}\!\left(\omega_0\,\|\,\omega_f\right)
\;=\; \frac{\delta\,\mathrm{Length}(\gamma_A)}{4G_N}
\;=\; \frac{c}{6}\,\delta\mathcal{A}\,,
\label{eq:main}
\end{equation}
where $\delta\mathcal{A} = \delta\,\mathrm{Length}(\gamma_A)/R$ is the dimensionless change in geodesic length for AdS radius $R$, and $c = 3R/(2G_N)$ is the Brown--Henneaux central charge~\cite{BrownHenneaux:1986}.


\paragraph*{Relative Entropy in the boundary CFT.}
In this section we review the derivation of the relative entropy between the vacuum state and a coherent excitation for a two-dimensional chiral CFT. For the sake of simplicity consider the local net of Weyl algebras $\mathsf{A}$ generated by the imaginary exponentials of the currents of a free massless scalar field $\phi$ on $\mathbb{M}^{1+1}$, writing $\mathsf{A}(\mathcal{O})$ for the von Neumann algebra of a region $\mathcal{O}$.  For a diamond $C_I$ with base $I\subset \Sigma=\{p\in\mathbb{M}^{1+1}\vert t=0\}$ of length $2\ell$ and centre $x^c$, the modular Hamiltonian $K^I$ for the pair $(\mathsf{A}(C_I),|\Omega\rangle)$, where $|\Omega\rangle$ denotes the vacuum vector, is given by~\cite[Eq~(2.23)]{Casini:2011kv}\cite{Longo:2020amm}\cite[Thm.~5.9.]{Longo:2021rag}\cite[Eq~(4)]{Cardy_2016}
\begin{equation}\label{eq:mod_diamond}
K^I \;=\; 2\pi\!\int_{-\infty}^{+\infty}\!
\frac{\ell^2-(x-x^c)^2}{2\ell}\,{:}T_{00}{:}(0,x)\,dx\,.
\end{equation}
Here ${:}T_{00}{:}(0,x)={:}(\partial_t\phi)^2{:}(0,x)-{:}(\partial_{x}\phi)^2{:}(0,x)$ denotes the energy density at $t=0$---automatically conserved and traceless in spacetime dimension $2$---understood in the sense of quadratic forms. We now compute the relative entropy for unitary excitations in the right chiral component of the algebra and make contact with the expressions derived in~\cite{Hollands:2019}. Let $u=t+x$ and $v=t-x$ be the null coordinates of $\mathbb{M}^{1+1}$. The field $\phi$ (and the associated currents) decomposes as the sum of two commuting chiral components $\phi(u,v)=\phi^{\operatorname{R}}(u)+\phi^{\operatorname{L}}(v)$. Given any domain $I^{\operatorname{R}}\times I^{\operatorname{L}}$, with $I^{\operatorname{R}/\operatorname{L}}$ subsets of the $u/v$ axes, the von Neumann algebra factorises,
\begin{equation}
    \mathsf{A}(I^{\operatorname{R}}\times I^{\operatorname{L}})= \mathsf{A}^{\operatorname{L}}(I^{\operatorname{L}})\otimes \mathsf{A}^{\operatorname{R}}(I^{\operatorname{R}})\,,
\end{equation}
where $\mathsf{A}^{\operatorname{R}/\operatorname{L}}$ is the net generated by the right/left chiral currents; for the diamond $C_I$, $I^{\operatorname{R}/\operatorname{L}}=[x^c-\ell,x^c+\ell]=I$. Correspondingly the energy density splits as
\begin{equation}
\begin{aligned}
    {:}T_{00}{:}(u,v)&={:}(\partial_u\phi^{\operatorname{R}})^2{:}(u)+{:}(\partial_v\phi^{\operatorname{L}})^2{:}(v)\\&=T^{\operatorname{R}}(u)+T^{\operatorname{L}}(v)\,.
\end{aligned}
\end{equation}
Consider now a unitary right excitation $U^{\operatorname{R}}(f)=e^{iT^{\operatorname{R}}(f)}$ with $f\in\mathcal{C}^\infty_0(\mathbb{R},\mathbb{R})$ and $\operatorname{supp}f\subset I$, so that $U^{\operatorname{R}}(f)\in\mathsf{A}^{\operatorname{R}}(I)\subset\mathsf{A}(C_I)$. The relative entropy between the vacuum state $\omega_0$ and the excited state $\omega_f$, implemented by $|\Psi_f\rangle := U^{\operatorname{R}}(f)|\Omega\rangle$ as a state on $\mathsf{A}(C_I)$, is
\begin{equation}
    S_{\operatorname{rel}}^I(\omega_0\Vert\omega_f)=\bra{\Psi_f} K^I \ket{\Psi_f}\,.
\end{equation}
On the constant-time slice $t=0$ we have $v=-u$, so that $T^{\operatorname{L}}$ is evaluated along the same Cauchy line as $T^{\operatorname{R}}$; setting $x^c=0$ without loss of generality and using that $T^{\operatorname{L}}$ has vanishing vacuum expectation value, this reduces to
\begin{equation}
  S^{I}_{\operatorname{rel}}(\omega_0\Vert\omega_f)= 2\pi\!\int_{-\infty}^{+\infty}\!
\frac{\ell^2-u^2}{2\ell}\, \bra{\Psi_f} \,T^{\operatorname{R}}(u) \ket{\Psi_f} \,du\,.
\end{equation}
This expression can be evaluated using Eq.~(2.13) of~\cite{Hollands:2019}, which is formulated in terms of the finite reparametrisation $\rho$ generated by $f$: the time-one map of the flow $\partial_s\rho_s(u)=f(\rho_s(u))$, $\rho_0(u)=u$. Since $T^{\operatorname{R}}$ has vanishing vacuum expectation value, its expectation value in the excited state is the Schwarzian anomaly of $\rho$~\cite{FEWSTER_2005}, and the relative entropy becomes (cf.~\cite[Eq.~(43)]{Hollands:2019})
\begin{equation}
\label{eq:Schwarzian-interval}
  S^{I}_{\operatorname{rel}}(\omega_0\Vert\omega_f) = -\frac{c}{12}
  \int_{-\ell}^{\ell}
  \frac{\ell^2-u^2}{2\ell}\,
  \bigl\{\rho(u),u\bigr\}\,du\,,
\end{equation}
where $\{\rho(u),u\}$ denotes the Schwarzian derivative and $c=1$ for the case of the massless field. Since ${\rho(u)=u}$ outside $\operatorname{supp}f$, where the flow acts trivially, the Schwarzian vanishes there and the integration is effectively confined to $I$. Equation \eqref{eq:Schwarzian-interval} holds for any two-dimensional chiral CFT, as proven in \cite{Panebianco_2020}.


\paragraph*{The Dual Bulk Geometry.} The geometry of AdS$_3$ spacetime is given, in global coordinates $(\tau,\rho,\varphi)$, by the metric \cite{Ryu:2006bv,Ammon_2015}
\begin{equation}
    g^{(0)} = R^2 (-\cosh^2\rho \, d\tau^2 + d\rho^2 + \sinh^2\rho \, d\varphi^2),
\end{equation}

\noindent where $R$ denotes the AdS radius corresponding to the constant negative curvature of the spacetime, and $\rho > 0$, with the conformal boundary located at $\rho \to \infty$.

Alternatively, in the Poincar\'e patch of AdS$_3$ (cf.~Refs.~\cite{BayonaBraga:2007pc,Ammon_2015}), the bulk metric can be written in coordinates $(t,x,z)$ as
\begin{equation}
g^{(0)}\;=\; \frac{R^2}{z^2}\bigl(-dt^2 + dx^2 + dz^2\bigr)\,,
\label{eq:AdS3}
\end{equation}
with $z>0$, which tends to zero towards the conformal boundary. As in the global coordinate system, a surface of constant Poincar\'e time $t=t_0$ geometrically corresponds to (a subset of) the hyperbolic plane $\mathbb{H}^2$ (cf.~Fig.~\ref{FigureAdS3}) with the metric
\begin{equation}
g^{(0)}\big|_{\mathbb{H}^2} \;=\; \frac{R^2}{z^2}\bigl(dx^2 + dz^2\bigr)\,.
\end{equation}
The vacuum geometry of AdS$_3$ is dual to the CFT vacuum. The coherent state $|\Psi_f\rangle$ is instead dual to a different asymptotically AdS$_3$ geometry. { Although ${|\Psi_f\rangle = U^{\operatorname{R}}(f)|\Omega\rangle}$ is generated by the stress tensor alone, it generically assigns non-vanishing expectation values to boundary operators other than $T$---dual to bulk matter fields---which would further affect the geometry in the bulk. However, in the pure gravity sector, where only $\langle T\rangle$ is retained and further matter contributions are neglected, the dual geometry is distinguished from the AdS$_3$ vacuum by its asymptotic stress-tensor data alone.} Transforming to lightlike coordinates $u = t+x$ and $v = t-x$, the Fefferman--Graham expansion of the bulk metric~\cite{FeffermanGraham1985, FeffermanGraham2007} yields the Ba\~nados metric 
\begin{equation}
\begin{aligned}
g^B &= \tfrac{R^2}{z^2}\,dz^2
- \Bigl(\tfrac{R^2}{z^2} + \tfrac{z^2}{R^2} L(u)\bar{L}(v)\Bigr)du\,dv \\
&\hspace{0.42cm} + L(u)\,du^2 + \bar{L}(v)\,dv^2\,.
\end{aligned}
\label{eq:Banados}
\end{equation}
as the most general solution to the Einstein equations that is asymptotically AdS$_3$, vacuum in the bulk but not on the boundary, and consistent with Brown--Henneaux boundary conditions~\cite{Banados:1998, Merbis:2019}. In particular, the coefficients $L(u)$ and $\bar{L}(v)$ appearing in this expansion are identified, via the holographic dictionary, with the expectation values of the chiral components of the boundary stress tensor,
\begin{equation}\label{Eq: B_coeff}
\langle T_{uu}\rangle = {\frac{c}{12\pi R^2}}\,L(u), \qquad
\langle T_{vv}\rangle = {\frac{c}{12\pi R^2} }\,\bar{L}(v),
\end{equation}
where $c = 3R/(2G_N)$ is the Brown--Henneaux central charge~\cite{BrownHenneaux:1986}. Note that for $L \equiv 0 \equiv \bar{L}$, the Ba\~nados metric reduces to the Poincar\'e metric~\eqref{eq:AdS3}, and in particular that it is locally isometric to the AdS$_3$ vacuum: it has no curvature singularities, no horizons, and no bulk matter.

If we further restrict ourselves to the right chiral sector (${\bar L=0}$), and linearize the Ba\~nados metric~\eqref{eq:Banados} about the AdS$_3$ vacuum, i.e.\
\begin{equation}
\label{eq:Banados-linear}
    g^B_{\mu\nu} = g^{(0)}_{\mu\nu} + h_{\mu\nu}\,,
\end{equation}
we find the explicit components
\begin{equation}
\begin{aligned}
h_{uu} &= L(u), \\
h_{uv}&=h_{vv}=h_{zz}=h_{zu}=h_{zv}=0\,.
\end{aligned}
\label{eq:h-components}
\end{equation}
of the metric perturbation. Note that in this specific case, the components of $h_{\mu\nu}$ are \emph{independent} of the coordinate $z$, which is one of the key facts underlying our main result given in Equation~\eqref{eq:main}. More precisely, this $z$-independence implies that $h_{uu}$ evaluated at the boundary $z=0$ and on the unperturbed RT geodesic $z=z_0(u)=\sqrt{\ell^2-(u-t_0)^2}$ gives the \emph{same} function of the coordinate $u$. On the constant-time slice $t = t_0 \equiv 0$, the RT geodesic of the unperturbed background $(h_{\mu\nu} = 0)$, anchored at the boundary points $u = \pm\ell$, is given by
\begin{equation}
    \gamma_A = \left\{ (u,z) \in \mathbb{H}^2 \;\middle|\; z = z_0(u),\;
    u \in [-\ell, \ell] \right\},
\end{equation}
where $z_0(u)$ is the semicircle
\begin{equation}
    z_0(u) = \sqrt{\ell^2 - u^2}, \qquad
    \frac{dz_0}{du} = \frac{-u}{\sqrt{\ell^2 - u^2}}.
\end{equation}
Accordingly, the induced metric on the geodesic \footnote{More precisely, we obtain the induced metric as $g^{(0)}_{\mathrm{ind}}=g^{(0)}(z=z_0(u),v=-u,u)$.} is given by the line element
\begin{equation}
g^{(0)}_{\mathrm{ind}} \;=\;
\Bigl(g^{(0)}_{zz}\bigl(\tfrac{dz}{du}\bigr)^{\!2} + g^{(0)}_{uu}\Bigr)\,du^2\,.
\end{equation}
In particular, given a perturbation of the form~\eqref{eq:Banados-linear}, the condition $h_{zz}=0$ implies that only $g_{uu}$ is perturbed along the geodesic. Writing the induced Ba\~nados metric coefficient as
\begin{equation}
g^{B}_{\mathrm{ind}}(u)
= g_{\mathrm{ind}}^{(0)}(u) + \delta g_{\mathrm{ind}}(u)\,,
\end{equation}
with $\delta g_{\mathrm{ind}}$ linear in $h_{\mu\nu}$, $z_0(u)$ thus gives
\begin{equation}
\delta g_{\mathrm{ind}}(u)
\;=\; h_{uu}\,\bigl(z_0(u),0,u\bigr)
\;=\; L(u)\,,
\label{eq:delta-g-ind}
\end{equation}
where the second equality follows from the $z$-independence of $h_{uu}$, as established in Equations~\eqref{eq:h-components}.

To obtain the corresponding first-order variation of the geodesic length of the RT curve $\gamma_A$, we first recall that
\begin{equation}
\begin{aligned}
    \mathrm{Length}(\gamma_A)& = \int_{-\ell}^{\ell}\sqrt{\vert g^{B}_{\mathrm{ind}}\vert}\,du
 \\&   =  \int_{-\ell}^{\ell}\sqrt{g_{\mathrm{ind}}^{(0)}}\,
  \sqrt{1 + \frac{\delta g_{\mathrm{ind}}}{g_{\mathrm{ind}}^{(0)}}}\,du
 \\& \simeq  \int_{-\ell}^{\ell}\sqrt{g_{\mathrm{ind}}^{(0)}}\left(
  1 + \frac{\delta g_{\mathrm{ind}}}{2\,g_{\mathrm{ind}}^{(0)}}\right)\,du,
\end{aligned}
\end{equation}
where we have kept only terms linear in $\delta g_{\mathrm{ind}}$ (cf.~\cite{PaulRoy2019}).

{ %
We now specify precisely which quantity this linearisation computes. Consider the one-parameter family of Ba\~nados metrics
\begin{equation}
g_\lambda \;=\; g^{(0)} + \lambda\, h\,,
\qquad h_{\mu\nu}\,du^\mu du^\nu = L(u)\,du^2\,,
\label{eq:banados-family}
\end{equation}
which interpolates between the AdS$_3$ vacuum at $\lambda=0$ and the Ba\~nados geometry, which is dual to $\ket{\Psi_f}$ { in the pure gravity sector described above}, at $\lambda=1$. Note that $h_{\mu\nu}$ is the \emph{exact} difference $g^B-g^{(0)}$, since in the right chiral sector $\bar L\equiv 0$ the metric~\eqref{eq:Banados-linear} is quadratic in no other component. Let
\begin{equation}
\gamma(\lambda) \;:=\; \mathrm{Length}\bigl[\gamma[g_\lambda]\bigr]
\label{eq:phi-def}
\end{equation}
denote the regulated length of the extremal geodesic anchored at $u=\pm\ell$ in the metric $g_\lambda$. The quantity we compute in~\eqref{eq:deltaLength} below is the first directional derivative
\begin{equation}
\delta\,\mathrm{Length}(\gamma_A)
\;=\; \gamma'(0)
\;=\; \frac{d}{d\lambda}\,\mathrm{Length}\bigl[\gamma[g^{(0)}+\lambda h]\bigr]\Big|_{\lambda=0}\,,
\label{eq:deltaLength-is-phiprime}
\end{equation}
i.e.\ the slope of $\gamma$ at the reference geometry, \emph{not} the total increment $\gamma(1)-\gamma(0)$. Because $\gamma[g^{(0)}]$ is extremal, the displacement of the curve contributes only at second order in $\lambda$, so $\gamma'(0)$ may be evaluated on the unperturbed geodesic $\gamma_0$, with $z_0(u)$ held fixed as in~\eqref{eq:delta-g-ind}--\eqref{eq:deltaLength}.

For $\mathrm{supp}\,f\Subset I$ the reparametrisation satisfies ${\rho(\pm\ell)=\pm\ell}$ and $\rho'(\pm\ell)=1$, leaving the endpoints and cutoff invariant, so the total length is unchanged, ${\gamma(1)=\gamma(0)}$. This reflects that $U^R(f)\in\mathcal{A}^R(I)$ acts within the interval algebra: the entanglement entropy of a unitary excitation of the vacuum does not change, $\Delta S_A=0$, and the Ryu--Takayanagi relation ${\Delta S_A=\bigl(\gamma(1)-\gamma(0)\bigr)/4G_N}$ holds here as $0=0$.

What the calculation returns is instead the nonvanishing slope~$\gamma'(0)$, evaluated on the exact source $L$; this is the quantity that enters the relative-entropy relation below. Explicitly, this yields}
\begin{equation}
\begin{aligned}
\delta\,\mathrm{Length}(\gamma_A)
&= \int_{-\ell}^{\ell}
  \bigl(\sqrt{g^B_{\mathrm{ind}}} - \sqrt{g_{\mathrm{ind}}^{(0)}}\bigr)\,du\\&\;=\; \int_{-\ell}^{\ell}\!
\frac{\delta g_{\mathrm{ind}}}{2\sqrt{\vert g_{\mathrm{ind}}^{(0)}\vert}}\,du \\
&\;=\; \frac{1}{R}\!\int_{-\ell}^{\ell}\!
\frac{\ell^2-u^2}{2\ell}\;
L(u)\,du\,,
\end{aligned}
\label{eq:deltaLength}
\end{equation}
where we used $\sqrt{\vert g_{\mathrm{ind}}^{(0)}\vert} = R\ell/(\ell^2-u^2)$. Substituting for $L(u)$, using Eq.~\eqref{Eq: B_coeff}, and using the relation $c=3R/(2G_N)$ for the Brown--Henneaux central charge, we obtain
\begin{equation}
\frac{\delta\,\mathrm{Length}(\gamma_A)}{4G_N}
\;=2\pi\;\int_{-\ell}^{\ell}\!
\frac{\ell^2-u^2}{2\ell}\,\langle T_{uu}\rangle(u)\,du\,.
\label{eq:deltaA-clean}
\end{equation}
Comparing with Expression~\eqref{eq:Schwarzian-interval} shows that also the relative entropy can be computed as the integral of the same function over the same interval $I$, and we find that the two expressions coincide.


\paragraph*{RT Formula for Relative Entropy.}
The equality of boundary relative entropy and bulk geodesic length rests on identifying the Ba\~nados source $L(u)\equiv L_\rho(u)$ with the boundary reparametrisation $\rho$, an identification we first assemble on the boundary and then promote to the bulk.

The Virasoro Ward identity, together with the vanishing of $\bra{\Omega}T\ket{\Omega}$, reduces the anomalous transformation under $u \mapsto \rho(u)$~\cite[Ch.~5]{DiFrancesco} (see also~\cite{FEWSTER_2005,Hollands:2019}) to the pure Schwarzian anomaly~\eqref{eq:Schwarzian-interval}, while the Fefferman--Graham dictionary~\eqref{Eq: B_coeff} identifies this same expectation value with the subleading coefficient of the metric expansion. Together they fix the chiral source,
\begin{equation}
    L_\rho(u) \;=\; -\tfrac{1}{2}\,\Sch{\rho(u)}{u}\,.
    \label{eq:LSchw}
\end{equation}

As it stands, however, \eqref{eq:LSchw} equates two pieces of \emph{boundary} data: the asymptotic source $L_\rho$ and the Schwarzian of $\rho$ agree as functions on the conformal boundary. In general this need not fix the bulk dual of $\ket{\Psi_f}$, since distinct bulk spacetimes may share the same conformal boundary and the same asymptotic stress tensor; the boundary agreement could be coincidental.

That it is not is guaranteed by the special rigidity of three-dimensional gravity. As noted above, AdS$_3$ gravity has no local (graviton) degrees of freedom, so the asymptotic data---the chiral source $L_\rho$---suffices to reconstruct the classical bulk geometry entirely~\cite{Banados:1998, Merbis:2019}. The boundary identification~\eqref{eq:LSchw} therefore cannot correspond to inequivalent bulk spacetimes: it fixes a single geometry. Roberts~\cite{Roberts_2012} makes this geometry explicit, exhibiting a bulk diffeomorphism $\Phi_\rho\colon\text{Poincar\'e-AdS}_3\to\text{Ba\~nados geometry}$ that realises in the bulk the reparametrisation $\rho=\mathrm{Exp}(f)$ generated by the same $f$ that defines $\ket{\Psi_f}$, mapping, to linear order in the metric perturbation, the AdS$_3$ vacuum to the Ba\~nados metric~\eqref{eq:Banados} with chiral source $L_\rho(u) = -\tfrac12\Sch{\rho(u)}{u}$. This promotes~\eqref{eq:LSchw} from a boundary identity to a statement about the dual geometry: { in the pure gravity sector,} the coherent state $\ket{\Psi_f} = U^R(f)\ket{\Omega}$ is dual to a Ba\~nados geometry with source $L_\rho$. Being a diffeomorphism, $\Phi_\rho$ produces no curvature, horizons, or bulk matter, consistent with our restriction to the right chiral sector $\bar L\equiv 0$.

\emph{Matching the two sides.}
With the bulk dual now fixed, both quantities of interest are the integral of the same chiral stress-tensor expectation value against the common interval weight $\beta(u):=(\ell^2-u^2)/2\ell$. On the boundary, the relative entropy~\eqref{eq:Schwarzian-interval} reads
\begin{equation}
 S^{I}_{\operatorname{rel}}(\omega_0\Vert\omega_f)
  \;=\;
  2\pi\!\int_{-\ell}^{\ell}\!
     \beta(u)\,\bra{\Psi_f} T^{\operatorname{R}}(u) \ket{\Psi_f}\,du\,,
\label{eq:Srel_modular}
\end{equation}
while the linearised geodesic-length formula~\eqref{eq:deltaA-clean}, derived using the $z$-independence of $h_{uu}$ from~\eqref{eq:h-components}, is the integral of the same quantity, $T^{\operatorname{R}}(u)=T_{uu}(u)$, against the same weight over the same interval,
\begin{equation}
\frac{\delta\,\mathrm{Length}(\gamma_A)}{4G_N}
\;=\; 2\pi\!\int_{-\ell}^{\ell}\!
\beta(u)\,\bra{\Psi_f} T_{uu}(u) \ket{\Psi_f}\,du\,.
\label{eq:RHS-final}
\end{equation}
Comparing~\eqref{eq:Srel_modular} and~\eqref{eq:RHS-final},
\begin{equation}
 S^{I}_{\operatorname{rel}}(\omega_0\Vert\omega_f)
  \;=\;
  \frac{\delta\,\mathrm{Length}(\gamma_A)}{4G_N}\,,
  \label{eq:main-derived}
\end{equation}
which establishes the main result of this letter, Eq.~\eqref{eq:main}. The relative entropy is thereby realised as a purely geometric functional of the asymptotic bulk data.


\paragraph*{Conclusion.}{ {
It is instructive to see where the relative entropy sits within the standard entropic decomposition. Formally, in QFT it obeys the first law
\begin{equation}
  S^{I}_{\operatorname{rel}}(\omega_0\Vert\omega_f)
  \;=\; \Delta\langle K\rangle - \Delta S_A\,,
  \label{eq:first-law}
\end{equation}
with $K$ the local modular Hamiltonian~\eqref{eq:mod_diamond} \footnote{As a quadratic form, $K$ coincides with the integral~\eqref{eq:mod_diamond} restricted to the finite interval.} and $\Delta S_A$ is the variation of the entanglement entropy.  The Ryu--Takayanagi relation connects the  entanglement-entropy variation with the change in extremal area which for the   example at hand is given by
\begin{equation}
  \Delta S_A
  \;=\; \frac{\gamma(1)-\gamma(0)}{4G_N}\,,
  \label{eq:RT-is-increment}
\end{equation}
so that the first law reads
\begin{equation}
  \boxed{\;
 S^{I}_{\operatorname{rel}}(\omega_0\Vert\omega_f)
  \;=\;  \Delta\langle K\rangle\;-\;\frac{\bigl[\gamma(1)-\gamma(0)\bigr]}{  4 G_N}
  \;}
  \label{eq:canonical-energy}
\end{equation}
whose right-hand side is the gravitational canonical energy of the Ba\~nados perturbation~\cite{Lashkari:2013,Faulkner:2013ica}.
Furthermore, if the excitation is a \emph{unitary} coherent excitation of the local algebra $\mathsf{A}^R(I)$~\cite{Hollands:2019,Longo_2019,Casini_2019,Panebianco_2020}, the Ryu--Takayanagi term~\eqref{eq:RT-is-increment} vanishes identically: $\Delta S_A=0$, equivalently $\gamma(1)-\gamma(0)=0$, since the extremal area is unchanged. The conventional entropy--geometry route, which reads geometric content off $\Delta S_A$, therefore has nothing to measure. The first law then forces the modular energy to equal the relative entropy, $\Delta\langle K\rangle = S^{I}_{\operatorname{rel}}(\omega_0\Vert\omega_f)$, which by our main result is the first-order length variation,
\begin{equation}
 \,\Delta\langle K\rangle \;=\;  \, S^{I}_{\operatorname{rel}}(\omega_0\Vert\omega_f) \;=\; \frac{\gamma'(0)}{4G_N}\,,
  \label{eq:conclusion-main}
\end{equation}
an exact geometric identity carried entirely by the modular energy.

We thus turn the standard correspondence around. Conventionally it is built around the entanglement entropy which, being ultraviolet-divergent for the type~III algebras of local QFT, carries geometric content only through its variation or under ad-hoc UV cut-offs---and only for states so close that the relative-entropy term is negligible. Our  Equation~\eqref{eq:conclusion-main} inverts the roles: we discard the  entanglement-entropy term (which for the unitary  {coherent} excitation case vanishes) and we focus on the finite one, namely the relative entropy. The Araki--Uhlmann relative entropy, that is finite with no regulator and no subtraction, becomes the primary object, and it is this finite quantity that fixes the geometry---recovering the Ryu--Takayanagi relation where that relation has content, and continuing to speak where the entanglement entropy is silent.

 We can as well recover a case studied in the literature~\cite{Blanco:2013}, namely the case of small perturbation of a state. This is done by expanding the formula for the Schwarzian derivative and take the first order. In particular, consider a family of  {small perturbations $\Omega_s$ ($s\in\mathbb{R}$) of $\Omega$.} Then, $S_{\text{rel}}(\Omega_s\Vert\Omega)$ is $0$ at first order in $s$ by  {positivity}. We recover this result by expanding $\left\{\rho_s(u),u\right\}$ in powers of $s$, i.e. ${\rho_s(u)=u+sf(u)+\mathcal{O}(s^2)}$, which implies
	\begin{equation}
	  \left\{\rho_s(u),u\right\}=s\,f'''(u)+\mathcal{O}(s^2).
	\end{equation}
	Since $\operatorname{supp}f\subset I$
		\begin{align}
		\nonumber S^{I}_{\operatorname{rel}}(\omega_f\Vert\omega_0) &= -s\frac{c}{12}
		\int_{-\ell}^{\ell}
		\frac{\ell^2-u^2}{2\ell}\, f'''(u)du\,+\mathcal{O}(s^2)\\&=0+\mathcal{O}(s^2),
        		\label{eq: Schwarzian-interval-pert}
	\end{align}
where in the last passage we used the fact that the boundary terms vanish, thus resulting in the vanishing relative entropy. }}
\paragraph*{Acknowledgements.}
The authors thank M. Fröb for discussing his results prior to publication, which are related to the present work.  L.S.  thanks N. Pinamonti and E. D'Angelo for various discussions on this topic. P.D. is grateful to L. Martinek for helpful discussions. { The authors thank P.~Roy for a careful reading of the manuscript and for pointing out an imprecision in the characterisation of the geodesic-length variation, which prompted the clarification in the paragraph following Eq.~\eqref{eq:deltaLength-is-phiprime}.} L.S. acknowledges financial support by Italian Ministry of University and
Research through the grant PRIN 2022ZE8SC4.
\bibliography{references}

\setcounter{equation}{0}
\setcounter{figure}{0}
\renewcommand{\theequation}{E.\arabic{equation}}
\renewcommand{\thefigure}{E.\arabic{figure}}

\end{document}